\documentclass[12pt]{article}

\usepackage{epsfig}

\title{Classical Electrodynamics:\\ A Tutorial on its
  Foundations\thanks{{\em Dedication to Erik W. Grafarend on the
      occasion of his 60th birthday:} Wir wissen, dass heutzutage auch
    die Geod\"aten relativistische Effekte bei ihren
    ``Triangulationen'' ber\"ucksichtigen m\"ussen. Wohl auch aus
    diesem Grunde hat Herr Grafarend immer ein offenes Ohr f\"ur
    entsprechende Theorien gehabt.  Die einfachste relativistische
    klassische Feldtheorie, die wir kennen, ist die Elektrodynamik.
    Wir widmen diese Ausarbeitung Herrn Grafarend zu seinem
    60.Geburtstage in der Hoffnung, dass er sich \"uber die sch\"onen
    Seiten dieser Darstellung genauso freut, wie wir es tun. Und dies
    umso mehr, als dass Herr Grafarend gleich am Anfang seiner
    Karriere sich intensiv mit Geometrie und Cartan-Formalismus
    auseinandergesetzt hat.}}

\author{Friedrich W.\ Hehl, Yuri N.\ Obukhov\thanks{Permanent address:
    Moscow State University.}$\,$, Guillermo F.\ Rubilar\\ Institute for 
  Theoretical Physics, University of Cologne\\ D-50923 K\"oln, Germany}

\date{27 July 1999}

\begin{document}

\maketitle 

\begin{abstract}
  We will display the fundamental structure of classical
  electrodynamics.  Starting from the axioms of (1) electric charge
  conservation, (2) the existence of a Lorentz force density, and (3)
  magnetic flux conservation, we will derive Maxwell's equations. They
  are expressed in terms of the field strengths $(E,{\cal B})$, the
  excitations $({\cal D},H)$, and the sources $(\rho,j)$.  This
  fundamental set of four microphysical equations has to be
  supplemented by somewhat less general constitutive assumptions in
  order to make it a fully determined system with a well-posed initial
  value problem. It is only at this stage that a distance concept
  (metric) is required for space-time. We will discuss one set of
  possible constitutive assumptions, namely ${\cal D}\sim E$ and
  $H\sim {\cal B}$. {\em file erik8a.tex, 1999-07-27}
\end{abstract}

\section{Introduction}

Is it worthwhile to reinvent classical electrodynamics after it has
been with us for more than a century? And after its quantized version,
quan\-tum electrodynamics (unified with the weak interaction) had
turned out to be one of the most accurately tested successful
theories? We believe that the answer should be affirmative. Moreover,
we believe that this reformulation should be done such that it is also
comprehensible and useful for experimental physicists and (electrical)
engineers\footnote{For this reason, we apply in this article the more
  widespread formalism of tensor analysis (``Ricci calculus'', see
  Schouten \cite{Schouten}) rather than that of exterior differential
  forms (``Cartan calculus'', see Frankel \cite{Ted}) which we basically
  prefer.}.

Let us collect some of the reasons in favor of such a reformulation.
First of all an ``axiomatics'' of electrodynamics should allow us to
make the fundamental structure of electrodynamics transparent, see,
e.g., Sommerfeld \cite{Sommerfeld} or \cite{Bopp,Honnef,Zirnbauer}. We
will follow the tradition of Kottler-Cartan-van Dantzig, see Truesdell
\& Toupin \cite{Truesdell} and Post \cite{Post}, and base our theory
on two experimentally well established axioms expressed in terms of
integrals, conservation of electric charge and magnetic flux, and a
local axiom, the existence of the Lorentz force.  All three axioms can
be formulated in a 4-dimensional (spacetime) continuum without using
the distance concept (i.e.\ without the use of a metric), see
Schr\"odinger \cite{Schroedinger}. Only the fourth axiom, a suitable
constitutive law, is specific for the ``material'' under consideration
which is interacting with the electromagnetic field. The vacuum is a
particular example of such a material. In the fourth axiom, the
distance concept eventually shows up and gives the 4-dimensional
continuum an additional structure.

Some of the questions one can answer with the help of such a general
framework are: Is the electric excitation $\cal D$ a {\em microscopic}
quantity like the field strength $E$? Is it justified to give $\cal D$
another dimension than $E$? The analogous questions can be posed for
the magnetic excitation $H$ and the field strength $\cal B$. Should we
expect a magnetic monopole and an explicit magnetic charge to arise in
such an electrodynamic framework? Can we immediately pinpoint the
(metric-independent) constitutive law for a 2-dimensional electron gas
in the theory of the quantum Hall effect? Does the non-linear
Born-Infeld electrodynamics fit into this general scheme? How do
Maxwell's equations look in an accelerated reference frame or in a
strong gravitational field as around a neutron star? How do they look
in a possible non-Riemannian spacetime?  Is a possible pseudoscalar
axion field compatible with electrodynamics? And eventually, on a more
formal level, is the calculus of exterior differential forms more
appropriate for describing electrodynamics than the 3-dimensional
Euclidean vector calculus and its 4-dimensional generalization? Can
the metric of spacetime be derived from suitable assumptions about the
constitutive law?

It is really the status of electrodynamics within the whole of physics
which comes much clearer into focus if one follows up such an
axiomatic approach.  

\section{Foliation of the 4-dimensional spacetime continuum} 

\begin{figure}[h]
\centering\epsfig{file=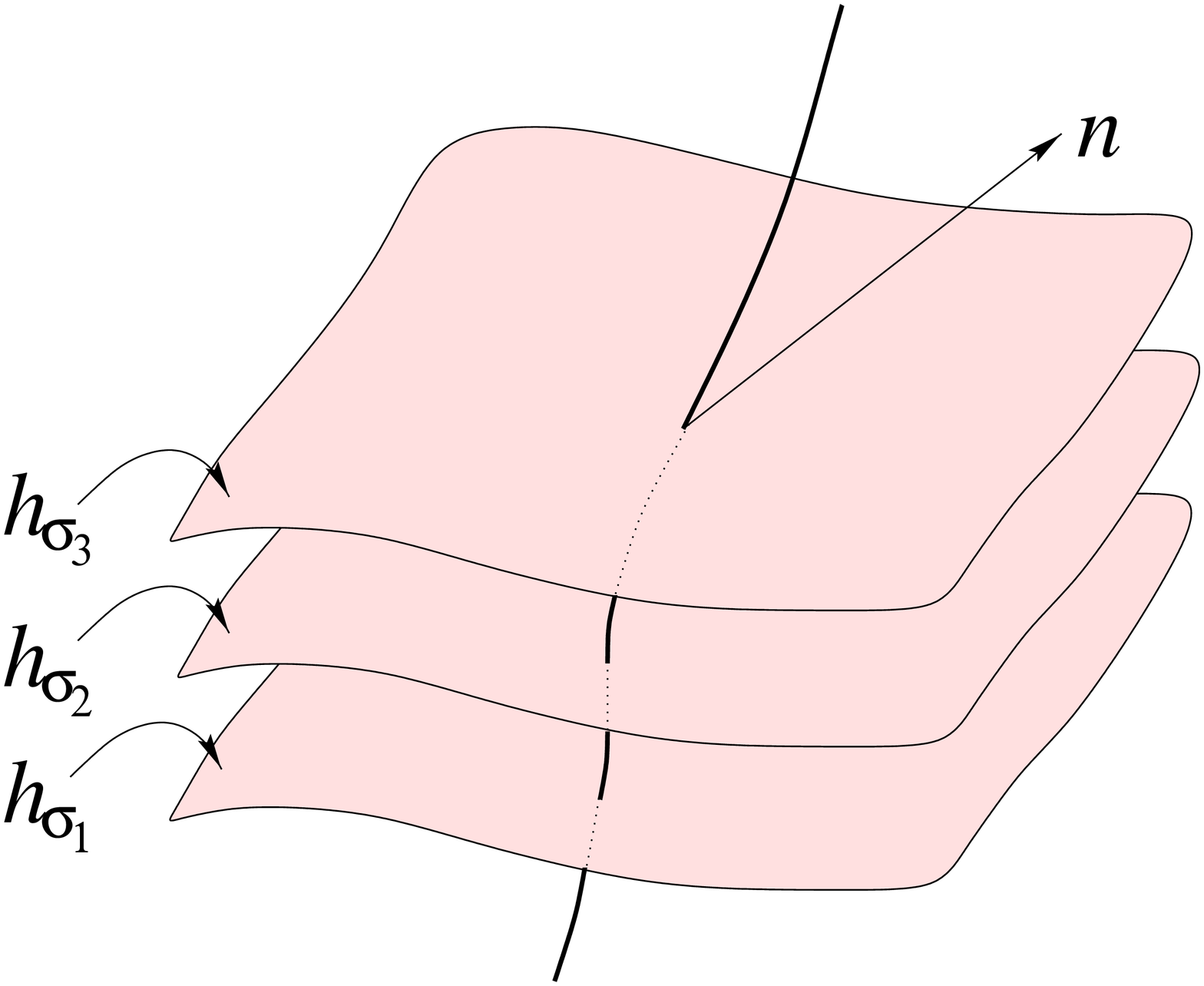, height=7cm, width=10cm} 
\caption{Foliation of spacetime: Each hypersurface $h_\sigma$ represents, 
   at a time $\sigma$, the 3-dimensional space of our perception one
   dimension of which is suppressed in the figure. The positive time
   direction runs upwards.}
\label{f1}
\end{figure}

{}From a modern relativistic point of view, the formulation of
electrodynamics has to take place in a 4-dimensional continuum
(differentiable manifold) which eventually is to be identified with
spacetime, i.e.\ with a continuum described by one ``time'' coordinate
$x^0$ and three ``space'' coordinates $x^1,x^2,x^3$ or, in short, by
coordinates $x^i$, with $i=0,1,2,3$. Let us suppress one space
dimension in order to be able to depict the 4-dimensional as a
3-dimensional continuum, as shown in Fig.\ref{f1}.

We assume that this continuum admits a foliation into a succession of
different leaves or hypersurfaces $h$. Accordingly, spacetime looks
like a pile of leaves which can be numbered by a monotonically
increasing (time) parameter $\sigma$. A leaf $h_\sigma$ is defined
by $\sigma(x^i)={const}$. It represents, at a certain time $\sigma$,
the ordinary 3-dimensional space surrounding us (in Fig.\ref{f1} it is
2-dimensional, since one dimension is suppressed).

At any given point in $h_\sigma$, we can introduce the covector
$k_i:=\partial_i\sigma$ and a 4-vector $n =(n^i)$ $= (n^0, n^1, n^2,
n^3)=(n^0, n^a)$ such that $n$ is normalized according to
\begin{equation}n^i\,k_i=n^i\partial_i\sigma=1\,. \end{equation}
 Here $a,b\dots=1,2,3$ and $i,j\dots=0,1,2,3$.
Furthermore, summation over repeated indices is always understood.
The vector $n^i$ is ``normal'' to the leaf $h_\sigma$, whereas the
covector $k_i$ is tangential\footnote{The term ``tangential'' is used
here in the sense of exterior calculus in which a covector (or 1-form)
is represented by two ordered parallel planes -- and the first plane
is tangential to $h_\sigma$.} to it.

With the pair $(n,k)$ we can construct projectors which decompose all
tensor quantities into {\em longitudinal}\hspace{0.3 mm} and {\em
  transversal}\hspace{0.3 mm} constituents with respect to the vector
$n$, see Fig.\ref{f2}. Indeed, the matrices
\begin{equation}
  L^i{}_j :=n^i\,k_j\quad{\rm and}\quad T^i{}_j := \delta^i_j -
  n^i\,k_j\quad{\rm with}\quad  L^i{}_j+T^i{}_j = \delta^i_j\,,
\end{equation} represent projection operators, i.e.
\begin{equation}
  L^i{}_j\,L^j{}_k =L^i{}_k\,, \quad T^i{}_j\,T^j{}_k =T^i{}_k\,,\quad
  L^i{}_j\,T^j{}_k = T^i{}_j\,L^j{}_k=0\,.
\end{equation} 
 
Taking an arbitrary covector $U_i$, we now can write it as
\begin{equation}
  U_i = {}^\bot\!U_i + \underline{U}_i\,,\quad{\rm where}\quad {}^\bot
  U_i := L^j{}_i\,U_j\quad{\rm and}\quad \underline{U}_i :=
  T^j{}_i\,U_j\,.
\end{equation}  Obviously
${}^\bot U_i$ describes the longitudinal component of the covector and
$\underline{U}_i$ its transversal component, with
$n^i\,\underline{U}_i =0$.  Analogously, for an arbitrary vector
$V^i$, we can write 
\begin{equation}
  V^i = {}^\bot\!V^i + \underline{V}^i\,,\quad{\rm where}\quad {}^\bot
  V^i := L^i{}_j\,V^j\quad{\rm and}\quad \underline{V}^i :=
  T^i{}_j\,V^j\,.
\end{equation}  
Its transversal component $\underline{V}^i$ fulfills
$\underline{V}^i\,k_i = 0$. This pattern can be straightforwardly
generalized to all tensorial quantities of spacetime.

For simplicity, we confine our attention to the particular case when
``adapt\-ed'' coordinates $x^i=(\sigma,x^a)$ are used and when the
``spatial'' components of $n$ vanish, i.e., $n^i=(1, 0, 0, 0)$. In that
case, we simply have $k_i = \partial_i\sigma = (1, 0, 0, 0)$ and hence
$\sigma$ can be treated as a formal ``time'' coordinate.

\section{Conservation of electric charge (axiom 1)}

The conservation of electric charge was already recognized as
fundamental law during the time of Franklin (around 1750) well before
Coulomb discovered the force law in 1785. Nowadays, at a time, at
which one can catch single electrons and single protons in traps and
can {\em count} them individually, we are more sure than ever that
electric charge conservation is a valid fundamental law of nature.
Therefore matter carries as a {\em primary quality} something called
electric charge which only occurs in positive or negative units of an
elementary charge $e$ (or, in the case of quarks, in $1/3$th of it)
and which can be counted in principle. Thus it is justified to
introduce the physical dimension of charge $q$ as a new and
independent concept.  Ideally one should measure a charge in units of
$e/3$. However, for practical reasons, the SI-unit C (Coulomb) is used
in laboratory physics.

\begin{figure}[h]
 \centering\epsfig{file=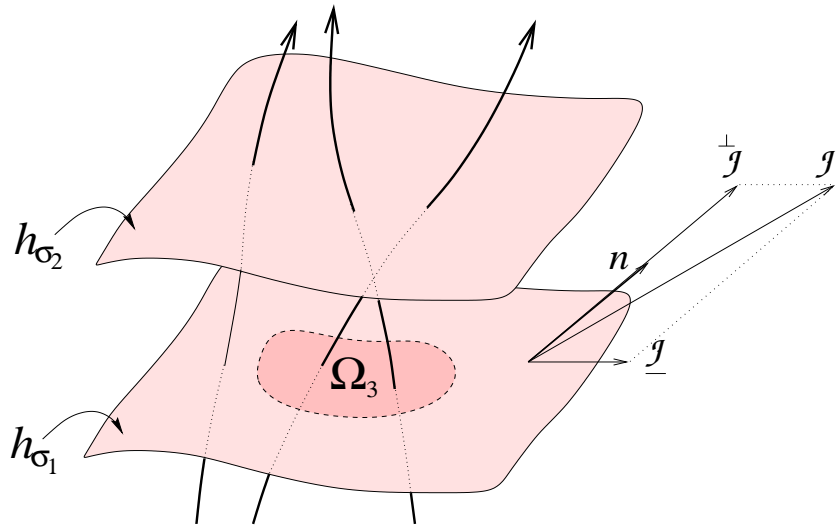, height=7cm, width=11cm}
 \caption{World lines, decomposition of the electric 
   current into the piece $^\bot{\cal J}$ longitudinal to $n$ and the
   transversal piece $\underline{{\cal J}}$, global conservation of
   charge.}
 \label{f2}
\end{figure}

Two remarks are in order: Charge is an additive (or extensive)
quantity that characterizes the source of the electromagnetic field.
It is prior to the notion of the electric field strength. Therefore it
is {\em not} reasonable to measure, as is done in the CGS-system of
units, the additive quantity charge in terms of the unit of force by
applying Coulomb's law. Coulomb's law has no direct relation to charge
conservation. Secondly, in the SI-system, for reasons of better
realization, the Ampere $A$ as current is chosen as the new
fundamental unit rather than the Coulomb. We have $C=A\,s$ ($s=$
second).

As a preliminary step, let us remind ourselves that, in a
4-dimen\-sio\-nal picture, the motion of a point particle is
described, as in Fig.\ref{f2}, by a curve in spacetime, by a so-called
worldline. The tangent vectors of this worldline represent the
4-velocity of the particle.

If we mark a 3-dimensional volume $\Omega_3$ which belong to a certain
hypersurface $h_\sigma$, then the total electric charge inside
$\Omega_3$ is
\begin{equation}\label{charge}
Q=\int\limits_{\Omega_3\subset h_\sigma}\rho\,dx^1dx^2dx^3\,,
\end{equation}
with $\rho$ as the electric charge density.  The total charge in
space, which we find by integration over the whole of space, i.e., by
letting $\Omega_3 \rightarrow h_\sigma$, is {\em globally} conserved.
Therefore the integral in (\ref{charge}) over each hypersurface
$h_{\sigma_1},h_{\sigma_2},\dots$ keeps the same value.

\begin{figure}[h]
 \centering\epsfig{file=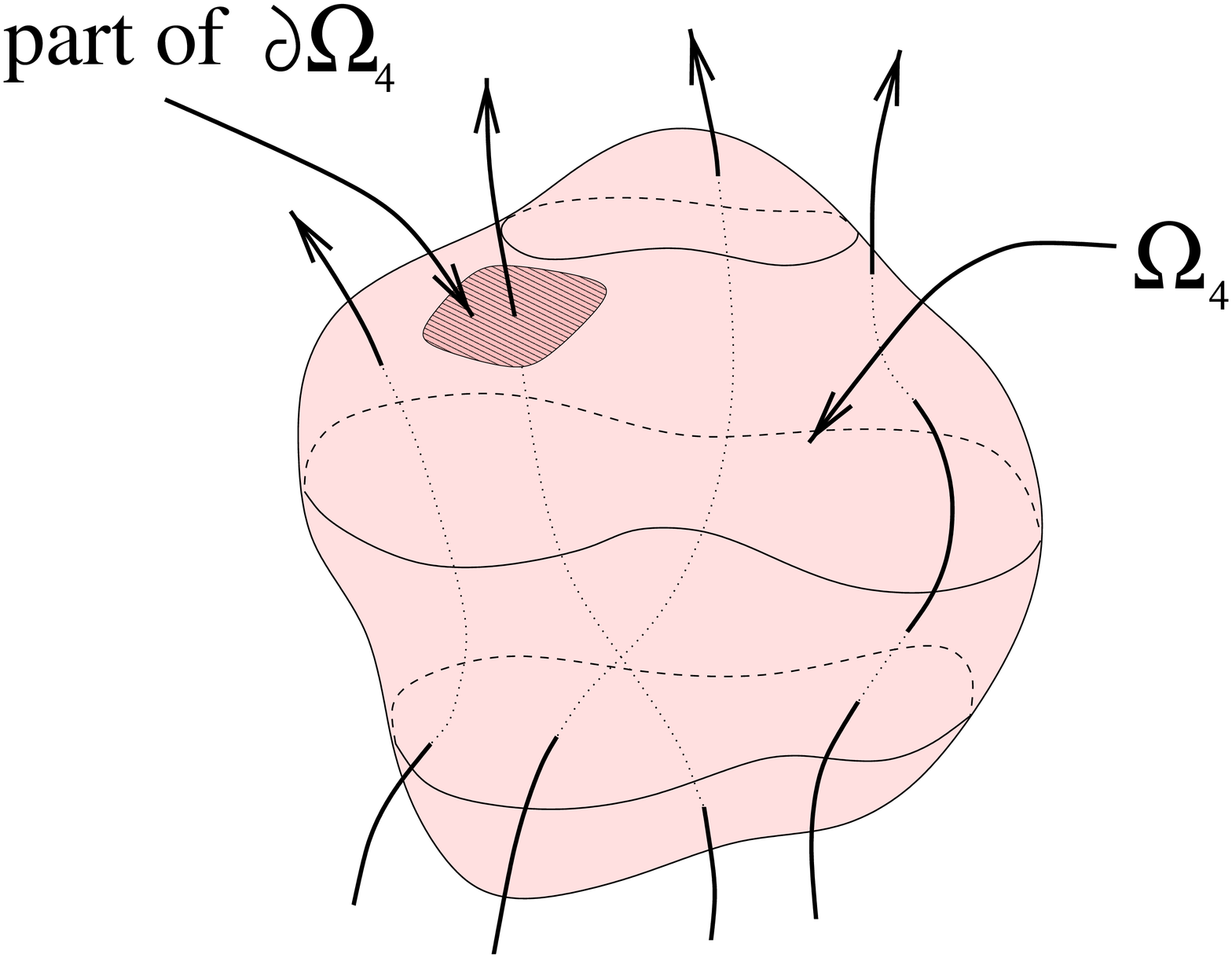, height=7cm, width=10cm}
 \caption{Local conservation of charge: Each worldline of a charged 
   particle that enters the finite 4-volume $\Omega_4$ via its
   boundary $\partial \Omega_4$ has also to leave $\Omega_4$.}
 \label{f3}
\end{figure}

The {\em local} conservation of charge, see Fig.\ref{f3}, translates
into the following fact: If a number of worldlines of particles with
one elementary charge enter a prescribed but arbitrary
4-dimen\-sion\-al volume $\Omega_4$, then, in classical physics, the
same number has to leave the volume.  If we count the entering
worldlines as negative and the leaving ones as positive (in conformity
with the direction of their normal vectors), then the (3-dimensional)
surface integral over the number of worldlines has to vanish.

Now, the natural extensive quantities to be integrated over a
3-dimen\-sional hypersurface are {\em vector densities,} see the
appendix.  Accordingly, in nature there should exist a 4-vector
density ${\cal J}^i$ with 4 independent components which measures the
charge piercing through an arbitrary 3-dimensional hypersurface.
Therefore, it generalizes in a consistent 4-dimensional formalism the
familiar concepts of charge density $\rho$ and current density $j^a$.
The axiom of local charge conservation then reads
\begin{equation}\label{axiom1}
  \int\limits_{\partial\Omega_4}{\cal J}^i\,d^{3}S_i=0\,,
\end{equation}
where the integral is taken over the (3-dimen\-sional) {\em boundary}
of an arbitrary 4-dimen\-sional volume of spacetime, with $d^{3}S_i$
being the 3-surface element, as defined in the appendix.

If we apply Stokes' theorem, then we can transform the 3-surface
integral in (\ref{axiom1}) into a 4-volume integral:
\begin{equation}\label{ax1stokes}
\int\limits_{\Omega_4}(\partial_i{\cal J}^i)\,d^{4}S=0\,.
\end{equation}
Since this is valid for an {\em arbitrary} 4-volume $\Omega_4$, we find the
local version of the charge conservation as
\begin{equation}\label{ax1local}
\partial_i {\cal J}^i=0\,.
\end{equation}
In this form, the law of conservation of charge is valid in arbitrary
coordinates. 

If one defines a particular foliation, then one can rewrite
(\ref{ax1local}) in terms of decomposed quantities that are
longitudinal and transversal to the corresponding normal vector $n$.
The 4-vector density ${\cal J}^i$ decomposes as
\begin{equation} \label{Jdecomp}
{\cal J}^i={}^\bot{\cal J}^i + \underline{\cal J}^i\,.
\end{equation}
When adapted coordinates are used, the decomposition procedure
simplifies and allows to define the 3-dimensional densities of charge
$\rho$ and of current $j^a$ as
\begin{equation} \rho:={}^\bot{\cal J}^0={\cal J}^0, \qquad j^a:=
\underline{\cal J}^a={\cal J}^a. \end{equation}
With this, one can rewrite the definition of charge (\ref{charge}) in
an explicitly coordinate invariant form
\begin{equation} \label{charge1}
  Q=\int\limits_{\Omega_3\subset h_\sigma} {\cal J}^i\,d^{3}S_i\,,
\end{equation}
since on $h_\sigma$ we have $d^{3}S_0=dx^1dx^2dx^3$ and $d^{3}S_a=0$.
Furthermore, Eq.(\ref{ax1local}) can be rewritten in (1+3)-form as the
more familiar continuity equation
\begin{equation}\label{continuity}
  \partial_\sigma\rho+\partial_aj^a=0\,.\end{equation}

The charge $Q$ in (\ref{charge1}) has the absolute
dimension\footnote{A theory of dimensions, which we are using, can be
  found in Post \cite{Post}, e.g.. A quantity has an {\em absolute}
  dimension, and if it is a density in spacetime we divide by
  $t\,l^3$.  The {\em components} pick up a $t$ (a $t^{-1}$) for an
  upper (a lower) temporal index and an $l$ (an $l^{-1}$) for an upper
  (a lower) spatial index. A statement, see \cite{Fitch}, that $E$ and
  $B$ must have the same dimension since they transform into each
  other is empty without specifying the underlying theory of
  dimensions.} $q$. The 4-current is a density in spacetime, and we
have $[{\cal J}]=q/(t\,l^3)$. Thus the components carry the dimensions
$[\rho]=[{\cal J}^0]=q/l^3 \stackrel{{\rm SI}}{=}C/m^3$ and
$[j^a]=[{\cal J}^a]=q/(t\,l^2)\stackrel{{\rm SI}}{=}A/m^2$.

\section{The inhomogeneous Maxwell equations as consequence}

Because of axiom 1 and according to a theorem of de Rham, see
\cite{Ted}, the electric current density from (\ref{axiom1}) or
(\ref{ax1local}) can be represented as a ``divergence'' of the
{\em electromagnetic excitation}:
\begin{equation}
  {\cal J}^i = \partial_j\,{\cal H}^{ij}\,,\qquad 
{\cal H}^{ij}=-{\cal H}^{ji}\,.\label{Jexact}
\end{equation}
The excitation ${\cal H}^{ij}$ is a contravariant antisymmetric tensor
density and has 6 independent components. One can verify that, due to
the antisymmetry of ${\cal H}^{ij}$, the conservation law is
automatically fulfilled, i.e., $\partial_i{\cal
  J}^i=\partial_i\partial_j{\cal H}^{ij}=0$.

The 4-dimensional set (\ref{Jexact}) represents the inhomogeneous
Maxwell equations. They surface here in a very natural way as a result
of charge conservation.  Charge conservation should not be looked at
as a consequence of the inhomogeneous Maxwell equations, but rather
the other way round, as shown in this tutorial. Of course, ${\cal
  H}^{ij}$ is not yet fully determined since 
\begin{equation}\label{Hgauge}
\tilde{{\cal H}}^{ij}={\cal H}^{ij}+\epsilon^{ijkl}\,\partial_k\psi_l
\end{equation}
also satisfies (\ref{Jexact}) for an arbitrary covector field
$\psi_i$.

The $(1+3)$-decomposition of ${\cal H}^{ij}$ is obtained similarly to
the decomposition of the current (\ref{Jdecomp}):
\begin{equation}\label{Hdecomp}
{\cal H}^{ij} = {}^{\bot}{\cal H}^{ij} + \underline{{\cal H}}^{ij}.
\end{equation}
The nontrivial components of the longitudinal and transversal parts
read
\begin{equation}
{\cal H}^{0a} = {}^{\bot}{\cal H}^{0a} = {\cal D}^a,\qquad 
{\cal H}^{ab} = \underline{{\cal H}}^{ab} = \epsilon^{abc}\,{H}_c\,,
\end{equation}
with the electric excitation ${\cal D}^a$ (historical name:
``dielectric displacement'') and the magnetic excitation ${H}_a$
(``magnetic field''). Here $\epsilon^{abc}$ is the totally antisymmetric
3-dimensional Levi-Civita tensor density with $\epsilon^{123}=1$.

If we substitute the decompositions (\ref{Jdecomp}) and
(\ref{Hdecomp}) into (\ref{Jexact}), we recover the 3-dimen\-sional
form of the inhomogeneous Maxwell equations,
\begin{equation}\label{3DMax} \partial_a {\cal D}^a = \rho, \qquad
  \epsilon^{abc}\partial_b H_c - \partial_\sigma {\cal D}^a
=
  j^a\,,
\end{equation} or, in symbolic notation, 
\begin{equation}\label{3DMaxsymb} 
  {\rm div}\, {\cal D} = \rho, \qquad {\rm curl}\,{{H}} -\dot{{\cal
      D}}= j \,.
\end{equation} Since electric charge
conservation is valid in microphysics, the corresponding Maxwell
equations (\ref{3DMax}) or (\ref{3DMaxsymb}) are also {\em
  microphysical} equations and with them the excitations ${\cal D}^a$
and ${H}_a$ are microphysical quantities likewise -- in contrast to
what is stated in most textbooks, see \cite{J} and \cite{Bergmann},
compare also \cite{Chambers}, e.g..

{}From (\ref{3DMax}) we can immediately read off $[{\cal
  D}^a]=[l\,\rho]=q/l^2\stackrel{\rm SI}{=}C/m^3$ and
$[H_a]=[l\,j^a]=q/(t\,l)\stackrel{\rm SI}{=}A/m$. Before we ever talked
about {\em forces} on charges, charge conservation alone gave us the
inhomogeneous Maxwell equations including the appropriate dimensions
for the excitations ${\cal D}^a$ and $H_a$.

Under the assumption that ${\cal D}^a$ vanishes inside an ideal
electric conductor, one can get rid of the indeterminacy of ${\cal
  D}^a$, as spelled out in (\ref{Hgauge}), and we can measure ${\cal
  D}^a$ by means of two identical conducting plates (``Maxwellian
double plates'') which touch each other and which are {\em separated}
in the ${\cal D}^a$-field to be measured. The charge on one plate is
then measured. Analogous remarks apply to $H_a$. Accordingly, the
excitations do have a direct {\em operational} significance.

\section{Force and field strengths (axiom 2)}

By now we have exhausted the information contained in the axiom 1 of
charge conservation. We have to introduce new concepts in order to
complete the fundamental structure of Maxwell's theory.  Whereas the
excitation ${\cal H}=({\cal D}^a,\,{H}_a)$ is linked to the charge
current ${\cal J}=(\rho,\,j^a)$, the electric and magnetic field
strengths are usually introduced as forces acting on unit charges at
rest or in motion, respectively. In the purely electric case with a
test charge $q$, we have in terms of components
\begin{equation}\label{Coul}
{F}_a\sim q\,E_a\,,\end{equation}
with ${F}$ as force and $E$ as electric field covector.

Let us take a (delta-function-like) test charge current ${\cal
  J}=(\rho,\,j^a)$ centered around a point with spatial coordinates
$x^a$. Generalizing (\ref{Coul}), the simplest relativistic ansatz for
defining the electromagnetic field reads:
\begin{equation}\label{fieldansatz1}
{\rm force\>\> density}\quad\sim\quad {\rm field\>\>strength}
  \times{\rm charge\>\> current\>\> density}\,.
\end{equation} 
We know from Lagrangian mechanics that the {\em force} $\sim\partial
L/\partial x^i$ is represented by a {\em covector} with the absolute
dimension of action $\hbar$ (here $\hbar$ is {\em not} the Planck
constant but rather only denotes its {\em dimension}). Accordingly, with the
covectorial force density $f_i$, the ansatz (\ref{fieldansatz1}) can
be made more precise as axiom 2:
\begin{equation}\label{fieldansatz2}
   f_i = F_{ij}\,{\cal J}^j\,,\qquad F_{ij}=-F_{ji}\,.
\end{equation}
The newly introduced covariant 2nd-rank 4-tensor $F_{ij}$ is the
electromagnetic field strength. The force density $f_i$ was postulated
to be normal to the current, $f_i\,{\cal J}^i=0$. Thus the antisymmetry of
the electromagnetic field strength is found, i.e., $F_{ij}$ depends on
6 independent components. We know the notion of force from mechanics,
the current density we know from axiom 1. Accordingly, axiom 2 is to
be understood as an {\em operational} definition of the
electromagnetic field strength $F_{ij}$.

With the decomposition
\begin{equation}\label{decoF}
F_{ij}={}^\bot F_{ij}+\underline{F}_{ij}\,,
\end{equation} 
we find the identifications for the electric field strength $E_a$ and
the magnetic field strength ${\cal B}^a$ (historical names: ``magnetic
induction'' or ``magnetic flux density''):
\begin{equation}\label{decompF}
  F_{a0} ={}^\bot F_{a0}=E_a,\qquad F_{ab} =\underline{F}_{ab} =
  \epsilon_{abc}\,{\cal B}^c.
\end{equation}
These identifications are reasonable since for the spatial components
of (\ref{fieldansatz2}) we recover the Lorentz force density and, for
the static case, Eq.(\ref{Coul}):
\begin{equation}
  f_a = F_{aj}\,{\cal J}^j=F_{a0}\,{\cal J}^0+F_{ab}\,{\cal
    J}^b=\rho\,E_a+ \epsilon_{abc}\,j^b\,{\cal B}^c\,.
\end{equation}
Symbolically, we have 
\begin{equation}
  f=\rho\,E+j\times{\cal B}\,.
\end{equation}
The time component of (\ref{fieldansatz2}) represents the
electromagnetic power density:
\begin{equation}
f_0=F_{0a}\,{\cal J}^a=-E_a\,j^a\,.
\end{equation}

\section{Conservation of magnetic flux (axiom 3)}

Axiom 2 on the Lorentz force gave us a new quantity, the
electromagnetic field strength with the dimension
$[F]=action/charge=\hbar/q=:\phi$, with $\phi=work\times
time/charge=voltage\times time\stackrel{\rm SI}{=}Vs=W\! b$. Here $W\!
b$ is the abbreviation for Weber. Thus its components carry the
following dimensions: $[E_a]=[F_{a0}]=\phi/(t\,l) \stackrel{\rm
  SI}{=}V/m$ and $[{\cal B}^c]=[F_{ab}]=\phi/l^2 \stackrel{\rm
  SI}{=}W\!b/m^2=T$ (for Tesla).

\begin{figure}[h]
 \centering\epsfig{file=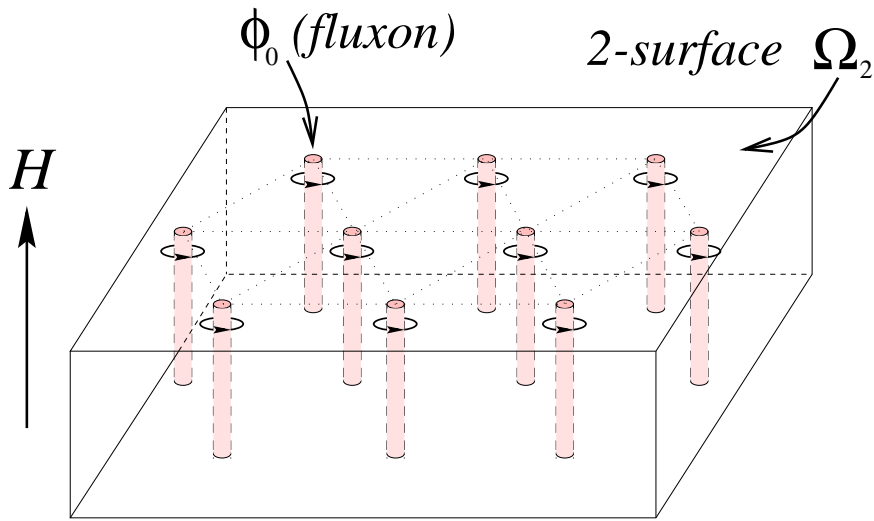, height=7cm, width=11.5cm}
 \caption{Sketch of an Abrikosov lattice in a type II superconductor in 
   3-di\-men\-sional space. In contrast to all the other figures, this is
   {\em not} an image of 4-dimensional spacetime.}
 \label{f4}
\end{figure}

We are in need of an experimentally established law that relates to
$F$. And we would prefer, as in the case of the electric charge, to
recur to a counting procedure. What else can we count in relation to
the electromagnetic field? Certainly {\em magnetic flux lines} in the
interior of a type II superconductor which is exposed to a
sufficiently strong magnetic field. And these flux lines are
quantized. In fact, they can order in a 2-dimensional triangular
Abrikosov lattice, see Fig.\ref{f4}. These flux lines carry a unit of
magnetic flux, a so-called flux quantum or {\em fluxon} with
$\Phi_0=h/(2e)=2.07\times 10^{-15} \,W\!b\,,$ see Tinkham
\cite{Tinkham}; here $h$ {\em is} the Planck constant and $e$ the
elementary charge. These flux lines can move, via its surface, in or
out of the superconductor, but they cannot vanish (unless two lines
with different sign collide) or spontaneously come into existence. In
other words, there is a strong experimental evidence that magnetic
flux is a conserved quantity.

The number $2$ in the relation $\Phi_0=h/(2e)$ is due to the fact that the
Cooper pairs in a superconductor consist of 2 electrons. Moreover,
outside a superconductor the magnetic flux is {\em not} quantized,
i.e., we cannot count the flux lines there with the same ease that we
could use inside. Nevertheless, as we shall see, experiments clearly
show that the magnetic flux is conserved also there.

As we can take from Fig.\ref{f4}, the magnetic flux should be defined
as a 2-dimensional spatial integral. These flux lines are additive and
we have
\begin{equation}\label{flux}
\Phi=\int\limits_{\Omega_2\subset h_\sigma} {\cal B}^a\,d^2S_a\,. 
\end{equation}

\begin{figure}[h]
 \centering\epsfig{file=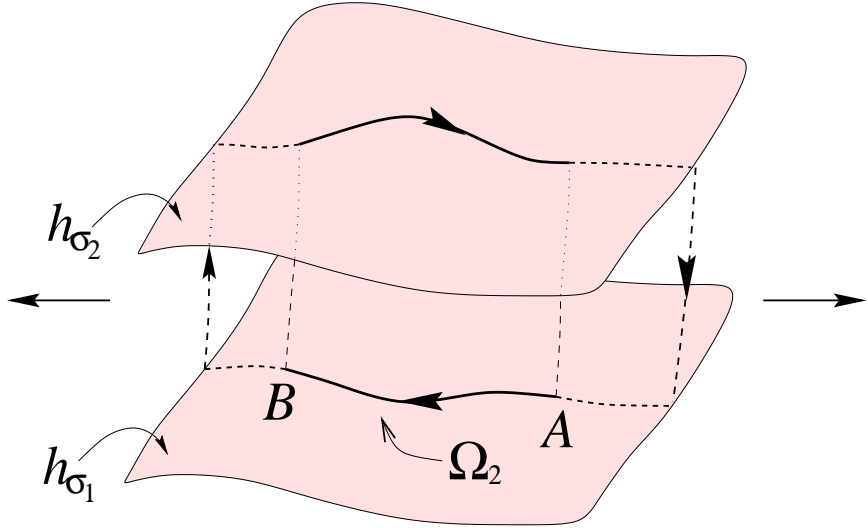, height=7cm, width=11.5cm}
 \caption{Magnetic flux in spacetime: The magnetic field ${\cal B}^a$, 
   if integrated over the interval $[A,B]$, represents, at time
   $\sigma_1$, the magnetic flux piercing through this 2-dimensional
   integration domain.}
 \label{f5}
\end{figure}

Here ${\cal B}^a$ is the magnetic field strength and $d^2S_a$ the
spatial 2-surface element. This definition of the magnetic flux should
be compared with the definition (\ref{charge}) of the charge.  Here,
in (\ref{flux}), we integrate only over $2$ dimensions rather than
over $3$ dimensions, as in the case of the charge in (\ref{charge}).
Thus in a spacetime picture in which one space dimension is
suppressed, see Fig.\ref{f5}, our magnetic flux integral looks like an
integral over a finite interval $[A,B]$ embedded into the hypersurface
$h_{\sigma_1}$.

Now we are going to argue again as in Sec.3. If $\Omega_2\rightarrow
\infty$, i.e., if we integrate over an infinite spatial 2-surface
($A\rightarrow +\infty,\,B\rightarrow -\infty$), then the total
magnetic flux at time $\sigma_1$ is given by (\ref{flux}).  If we
propagate that interval into the (coordinate) future, see the interval
on the hypersurface $h_{\sigma_2}$, then magnetic flux conservation
requires the constancy of the integral $\Phi$. In other words, if we
orient the integration domain suitably, the loop integral, the domain
of which is drawn in Fig.\ref{f5}, has to vanish since no flux is
supposed to leak out along the dotted ``vertical'' domains at spatial
infinity.

\begin{figure}[h]
 \centering\epsfig{file=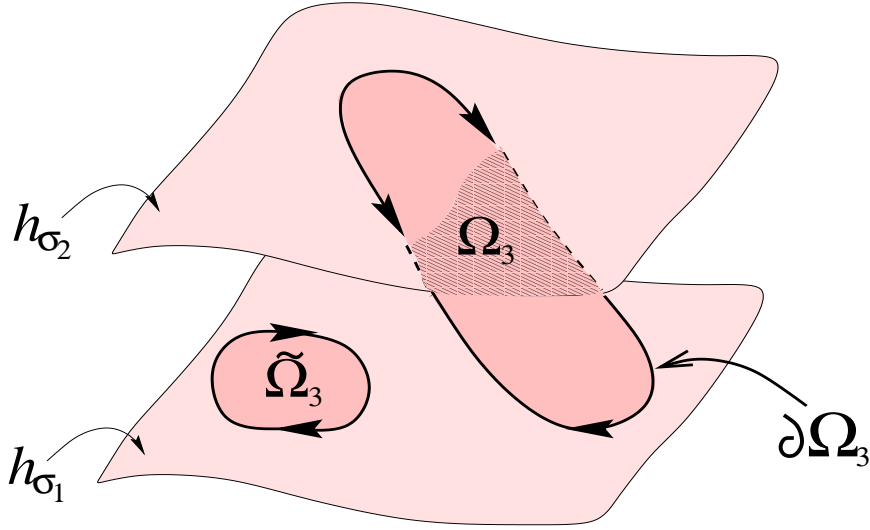, height=7cm, width=11.5cm}
 \caption{Conservation of magnetic flux in spacetime: Consider the 
   arbitrary 3-dimensional integration domain $\Omega_3$.  The
   integral vanishes of the field strength $F_{ij}$ over the
   2-dimensional boundary $\partial\Omega_3$ of the 3-dimensional
   domain $\Omega_3$. The analogous is true for the flux integral over
   $\partial\tilde{\Omega}_3$.}
 \label{f6}
\end{figure}

Analogously as we did in the case of charge conservation, we want to
formulate a corresponding local conservation law in an explicitly
covariant way. We saw that the global conservation of magnetic flux is
expressed as the vanishing of the integral of ${\cal B}$ over the
particular 2-dimensional loop in Fig.\ref{f5}. In a 4-dimensional
covariant formalism, the natural intensive objects to be integrated
over a 2-dimensional region are second order antisymmetric covariant
tensors, see the appendix. The magnetic field strength ${\cal B}$ is
just a piece of the electromagnetic field strength $F$.  Thus, it is
clear that the natural local generalization of the magnetic flux
conservation, our axiom 3, is
\begin{equation}\label{axiom3}
\int\limits_{\partial\Omega_3}\frac {1}{2}\,F_{ij}\,
d^2S^{ij} = 
0\,,
\end{equation}
where the integral is taken over the boundary of an arbitrary
3-dimensional hypersurface of spacetime, as is sketched in
Fig.\ref{f6}. We apply Stokes' theorem
\begin{equation}\label{axiom3'}
  \int\limits_{\Omega_3} \epsilon^{ijkl}\partial_{[j}\,F_{kl]}\, 
d^3S_i = 0\,,
\end{equation}
and, since the volume is arbitrary, we have the local version of
magnetic flux conservation as 
\begin{equation}\label{axiom3'''}
\partial_{[i}F_{jk]} = 0.
\end{equation}

We substitute the decomposition (\ref{decoF}) into
(\ref{axiom3'''}). Then we find the homogeneous Maxwell equations,
\begin{equation}\label{homMax}
  \partial_a {\cal
    B}^a=0\,,\qquad\epsilon^{abc}\partial_bE_c+\partial_\sigma{\cal
    B}^a=0\,
\end{equation} or, symbolically, 
\begin{equation}\label{homMax'}
  {\rm div}\, {\cal B}=0\,,\qquad {\rm curl}\, E+\dot{{\cal B}}=0\,.
\end{equation}
Thus both, the sourcelessness of ${\cal B}^a$ and the Faraday
induction law follow from magnetic flux conservation. Both laws are
experimentally very well verified and, in turn, strongly support the
axiom of the conservation of the magnetic flux.

The recognition that Maxwell's theory, besides on charge conservation,
is based on magnetic flux conservation, sheds new light on the
possible existence of magnetic monopoles. First of all, careful search
for them has not lead to any signature of their possible existence,
see \cite{He}. Furthermore, magnetic flux conservation would be
violated if we postulated the existence of a current on the right hand
side of (\ref{axiom3'''}). Now, Eq.(\ref{ax1local}) is the analog of
(\ref{axiom3'''}), at least in our axiomatic set-up. Why should we
believe in charge conservation any longer if we gave up magnetic flux
conservation? Accordingly, we assume -- in contrast to most elementary
particle physicists, see Cheng \& Li \cite{ChengLi} -- that in
Maxwell's theory proper there is no place for a magnetic
current\footnote{This argument does not exclude that, for {\em
    topological} reasons, the integral in (\ref{axiom3}) could be
  non-vanishing, as in the case of a Dirac monopole with a string, see
  \cite{Marc}.} on the right hand side of (\ref{axiom3'''}).

\section{Constitutive law (axiom 4)}\label{mat}

The Maxwell equations (\ref{Jexact}) and (\ref{axiom3'''}) or, in the
decomposed version, (\ref{3DMaxsymb}) and (\ref{homMax'}),
respectively, encompass altogether 6 partial differential equations
with a first order time derivative (the 2 remaining equations can be
understood as constraints to the initial configuration). Since
excitations and field strengths add up to $6+6=12$ independent
components, certainly the Maxwellian set is underdetermined with
respect to the time propagation of the electromagnetic field. What we
clearly need is a relation between the excitations and the field
strengths. As we will see, these so-called constitutive equations
require additional knowledge about the properties of spacetime whereas
the Maxwell equations, as derived so far, are of universal validity as
long as classical physics is a valid approximation. In particular, in
the Riemannian space of Einstein's gravitational theory the Maxwell
equations look just the same as in (\ref{Jexact}) and
(\ref{axiom3'''}). There is no adaptation needed of any kind, see
\cite{PLH}.

If we investigate macroscopic matter, one has to derive from the
microscopic Maxwell equations by statistical procedures the {\em
  macroscopic} Maxwell equations. They are expected to have the same
structure as the microscopic ones. But let us stay, for the time
being, on the microscopic level. 

Then we can make an attempt with a {\em linear} constitutive relation
between ${\cal H}^{ij}$ and $F_{kl}$,
\begin{equation}\label{const}
  {\cal H}^{ij}=\frac{1}{2}\,\tilde{\chi}^{ijkl}\,F_{kl}=\frac{1}{2}\,f
    {\chi}^{ijkl}\,F_{kl}\,,
\end{equation} 
with the tensor density ${\chi}^{ijkl}$ that is characteristic
for the spacetime under consideration.  We require $[{\chi}]=1$,
i.e., for the dimensionfull scalar factor factor $f$ we have
$[f]=q/\phi=q^2/h\stackrel{\rm SI}{=}C/(Vs)=A/V=1/\Omega$. The
dimensionless ``modulus'' ${\chi}^{ijkl}$, because of the
antisymmetries of ${\cal H}^{ij}$ and $F_{kl}$, obeys
\begin{equation}\label{const'}
{\chi}^{ijkl}=-{\chi}^{jikl}=-{\chi}^{ijlk}\,.
\end{equation}
Moreover, if we assume the existence of a Lagrangian density for the
electromagnetic field ${\cal L}\sim {\cal H}^{ij}F_{ij}$, then we have
additionally the symmetries
\begin{equation}\label{const''}
{\chi}^{ijkl}={\chi}^{klij}\,,\qquad{\chi}^{[ijkl]}=0\,.
\end{equation}
The vanishing of the totally antisymmetric part comes about since the
corresponding Euler-Lagrange derivative of ${\cal L}$ with respect to
the 4-potential $A_i$ identically vanishes; here
$F_{ij}=2\partial_{[i}A_{j]}$. For ${\chi}^{ijkl}$, this leaves 20
independent components\footnote{With such a linear constitutive law it
  is even possible to {\em derive}, up to a con\-for\-mal factor, a
  metric of spacetime, provided one makes one additional assumption,
  see \cite{Schoenberg,OH}.}. One can take such moduli, if applied on
a macrophysical scale, for describing the electromagnetic properties
of anisotropic crystals, e.g.. Then also non-linear (for
ferromagnetism) and spatially non-local constitutive laws are in use.

The simplest linear law is expected to be valid in vacuum.
Classically, the vacuum of spacetime is described by its metric tensor
$g^{ij}=g^{ji}$ that determines the temporal and spatial distances of
neighboring events. Considering the symmetry properties of the density
${\chi}^{ijkl}$, the only ansatz possible, up to an
arbitrary constant, seems to be
\begin{equation}\label{vac}
  {\chi}^{ijkl}=\sqrt{-\det
    g_{mn}}\,\left(g^{ik}g^{jl}-g^{jk}g^{il}\right)\,.
\end{equation}
Note that ${\chi}^{ijkl}$ is invariant under a rescaling of the
metric $g_{ij}\rightarrow \Omega^2 g_{ij}$, with an arbitrary function
$\Omega(x^i)$. Using this freedom, we can always normalize the
determinant of the metric to 1.

As an example, let us consider the {\em flat} spacetime metric of a
Minkowski space in Minkow\-skian coordinates,
\begin{equation}\label{Mink}
\eta^{ij} = {{\sqrt{c}}}\left(\begin{array}{crrr}c^{-2} & 0 & 0 & 0 
\\ 0 & -1 & 0 & 0\\ 0& 0 & -1 & 0\\ 0& 0& 0& -1\end{array}\right).
\end{equation}
If we substitute (\ref{Mink}) into (\ref{vac}) and, in turn,
Eq.(\ref{vac}) and $f=\sqrt{\varepsilon_0/\mu_0}$ into (\ref{const}),
then we eventually find the well-known vacuum (``Lorentz aether'')
relations,
\begin{equation}
  {\cal H}^{ij}=\sqrt{\frac{\varepsilon_0}{\mu_0}}
  \eta^{ik}\eta^{jl}\,F_{kl} \qquad{\rm or}\qquad {\cal
    D}=\varepsilon_0\,E\,,\quad H=(1/\mu_0)\,{\cal B}\,.\label{aether}
\end{equation}
The law (\ref{aether}) converts Maxwell's equations, for vacuum, into
a system of differential equations with a well-determined initial
value problem.
\bigskip

\noindent {\em Acknowledgments:} We are grateful to Marc Toussaint for
discussions on magnetic monopoles. G.F.R.\ would like to thank the
German Academic Exchange Service DAAD for a graduate fellowship
(Kennziffer A/98/00829). 

\appendix

\section{Four-dimensional calculus without metric and integrals}

In a 4-dimensional space, in which arbitrary coordinates $x^i$ are
used, with $i=0,1,2,3$, one can define derivatives and integrals of
suitable antisymmetric covariant {\em tensors} and antisymmetric
contravariant {\em tensor densities} without the need of a metric. The
tensors are used for representing intensive quantities (how strong?),
the tensor densities for extensive (additive) quantities (how much?).
The natural formalism for defining integrals in a coordinate invariant
way is exterior calculus, see Frankel \cite{Ted}. However, we will use
here tensor calculus, see Schouten \cite{Schouten} and also
Schr\"odinger \cite{Schroedinger}, which is more widely known under
physicists and engineers.  \bigskip

\noindent{\bf Integration over 4-dimensional regions -- scalar densities}
\medskip

Consider a certain 4-dimensional region $\Omega_4$. Then a 
integral over $\Omega_4$ is of the form
\begin{equation}\label{A1}\int_{\Omega_4}{\cal A} \, d^4S\, , 
\end{equation}
where $d^4S:=dx^0 dx^1 dx^2 dx^3$ is the 4-volume element which is a
scalar density of weight $-1$. We want this integral to be a scalar,
i.e., that its value does not depend on the particular coordinates we
use. Then the integrand ${\cal A}$ has to be a scalar density of
weight $+1$. In other words, when using the tensor formalism, the
natural quantity required to formulate an invariant integral over a
4-dimensional region is a scalar density of weight $+1$.  \bigskip

\noindent{\bf Integration over 3-dimensional regions -- vector densities}
\medskip

 Now we want to define invariant integrals over some 3-dimensional 
hypersurface $\Omega_3$ in a four-dimensional space which can be defined by 
the parameterization $x^i=x^i(y^a)$, $a,b,c=1,2,3$, where $y^a$ are also 
arbitrary coordinates on $\Omega_3$. Then we call 
\begin{equation}\label{A2} d^3S_i:=\frac{1}{3!}\, \epsilon_{ijkl}\, 
  \frac{\partial x^j}{\partial y^a} \frac{\partial x^k}{\partial y^b}
  \frac{\partial x^l}{\partial y^c}\, \epsilon^{abc}\, dy^1 dy^2 dy^3
\end{equation}
the {\em 3-surface element on $\Omega_3$}. This quantity is constructed 
by using only objects that can be defined in a general 4-dimensional space 
without metric or connection. It can be constructed as soon as we specify 
the parameterization of $\Omega_3$. 
Here $\epsilon_{ijkl}$ is the 4-dimensional Levi-Civita 
tensor density of weight $-1$ and 
$\epsilon^{abc}$ the 3-dimensional Levi-Civita tensor 
density of weight 
$+1$ on $\Omega_3$. Furthermore, this hypersurface element turns out to be a 
covector density of weight $-1$ with respect to 4-dimensional coordinate 
transformation. With this integration element to our 
disposal, the natural form of an invariant integral over $\Omega_3$ is
\begin{equation}\label{A3} \int_{\Omega_3}{\cal A}^i d^3S_i\,. \end{equation}
Therefore, the natural object to be integrated over $\Omega_3$ in
order to obtain an invariant result is a vector density of weight
$+1$.  \bigskip

\noindent{\bf Integration over 2-dimensional regions -- covariant 
tensors or contravariant tensor densities}
\medskip

Analogously, we can parameterize a 2-dimensional region $\Omega_2$ by
means of $x^i=x^i(z^\alpha)$, $\alpha,\beta=1,2$, where $z^\alpha$ are
arbitrary coordinates on $\Omega_2$. Then we can immediately construct
the following 2-surface element
\begin{equation}\label{A4} d^2S_{ij}:=\frac{1}{2}\,\epsilon_{ijkl}\, 
  \frac{\partial x^k}{\partial z^\alpha} \frac{\partial x^l}{\partial
    z^\beta} \,\epsilon^{\alpha\beta}\, dz^1 dz^2, \end{equation}
where $\epsilon^{\alpha\beta}$ is the Levi-Civita density of weight
$+1$ on $\Omega_2$. This surface element is an antisymmetric second
order covariant tensor density of weight $-1$.  Then an invariant
integral is naturally defined as
\begin{equation}\label{A5} \int_{\Omega_2}\frac{1}{2}{\cal A}^{ij}\, 
  d^2S_{ij}\, ,\end{equation}
with ${\cal A}^{ij}$ being an antisymmetric second order contravariant
tensor density of weight $+1$.

Alternatively, one can write the same integral in terms on an
antisymmetric second order covariant tensor
$A_{ij}:=\frac{1}{2}\epsilon_{ijkl}{\cal A}^{kl}$ and an antisymmetric
second order contravariant surface element 
\begin{equation}\label{A6}d^2S^{ij}:=\frac{\partial
    x^k}{\partial z^\alpha} \frac{\partial x^l}{\partial
    z^\beta}\epsilon^{\alpha\beta}\, dz^1 dz^2\,,\end{equation} such
that
\begin{equation}\label{A7}  \int_{\Omega_2}\frac{1}{2}{\cal A}^{ij}\, 
  d^2S_{ij}= \int_{\Omega_2}\frac{1}{2} A_{ij}\, d^2S^{ij}\, .
\end{equation} 
Since extensive quantities are represented by densities, we would take
the first integral for them, whereas for intensive quantities the
second integral should be used. Analogous considerations can be applied to 
(\ref{A1}) and (\ref{A3}).\bigskip

\noindent{\bf Stokes' theorem}\medskip 

Stokes' theorem gives us as particular cases the following integral
identities (see \cite{Schouten} p.67 et seq.):
 \begin{equation} \int_{\Omega_4}(\partial_i {\cal J}^i)\,
   d^4S =\int_{\partial \Omega_4}{\cal J}^i \, d^3S_i\,,
 \end{equation}
\begin{equation} \int_{\Omega_3}(\partial_j {\cal H}^{ij})\, d^3S_i = 
  \int_{\partial \Omega_3}\frac{1}{2}\, {\cal H}^{ij} \, d^2S_{ij}\,.
\end{equation}

\section{Decomposition of totally antisymmetric tensors 
  into longitudinal and trans\-versal pieces}

Here we provide the decomposition formulas for totally antisymmetric
covariant and contravariant tensors, which are the natural
generalization of the decomposition of vectors and covectors.

We start by considering an antisymmetric covariant tensor of rank $p$,
namely $U_{i_1\dots i_p}$. Its longitudinal and transversal components
are given by
\begin{equation} 
{}^\bot U_{i_1\dots i_p}=p\ L^m_{\ [i_1|} U_{m|i_2\dots i_p]}\,, 
 \qquad \underline{U}_{i_1\dots i_p}=(p+1) L^m_{\ [m} U_{i_1\dots i_p]}\,, 
\end{equation}
where $k_i:=\partial_i\sigma$, and $L^i_{\ j}:=n^i k_j$. 
They fulfill the following properties: 
\begin{equation} n^{i_1}\underline{U}_{i_1\dots i_p}=0\,, \qquad
  n^{i_1}{}^\bot U_{i_1\dots i_p}=n^{i_1} U_{i_1\dots i_p}\,.
\end{equation} For $p=1,2,3,4$ we can explicitly write:
\begin{center}
\begin{tabular}{|c|c|c|c|}
\hline $p$ & quantity & definition & explicitly \\ \hline
\hline $1$ & ${}^\bot U_i$ & $L^m_{\ i}U_m$ & $n^m k_i U_m$ \\ 
\hline $2$ & ${}^\bot U_{ij}$ & $2L^m_{\ [i|} U_{m|j]}$ & 
       $n^m\left(k_{i} U_{mj}-k_{j} U_{mi}\right)$\\ 
\hline $3$ & ${}^\bot U_{ijk}$ & $3L^m_{\ [i|} U_{m|jk]}$ &
       $n^m\left(k_{i} U_{mjk}+
       k_{j} U_{mki}+k_{k} U_{mij}\right)$ \\
\hline $4$ & ${}^\bot U_{ijkl}$ & $4L^m_{\ [i|} U_{m|jkl]}$ & 
       $n^m\left(k_{i} U_{mjkl}-
       k_{j} U_{mkli}+ k_{k} U_{mlij}-
       k_{l} U_{mijk}\right)$ \\ 
\hline
\end{tabular}
\end{center}
Now we turn to $V^{i_1\dots i_p}$, an antisymmetric contravariant
tensors of rank $p$. We define the decomposition as
\begin{equation} {}^\bot V^{i_1\dots i_p}=p\ L^{[i_1|}_{\ \ m} V^{m|i_2 
\dots i_p]}\,, 
\qquad \underline{V}^{i_1\dots i_p}=(p+1) L^{[m}_{\ \ m} V^{i_1\dots i_p]}\,. 
\end{equation}
They fulfill
\begin{equation} k_{i_1}\underline{V}^{i_1\dots i_p}=0\,, \qquad
    k_{i_1}{}^\bot V^{i_1\dots i_p}= k_{i_1} V^{i_1\dots i_p}\,. \end{equation}
 For $p=1,2,3,4$ we have the following explicit expressions for the 
longitudinal components:
\begin{center}
\begin{tabular}{|c|c|c|c|}
\hline $p$ & quantity & definition & explicitly \\ \hline
\hline $1$ & ${}^\bot V^i$ & $L^i_{\ m} V^m$ & $n^i k_m V^m$ \\ 
\hline $2$ & ${}^\bot V^{ij}$ & $2L^{[i|}_{\ \ m} V^{m|j]}$ & 
       $k_m\left(n^i V^{mj}-n^j V^{mi}\right)$\\ 
\hline $3$ & ${}^\bot V^{ijk}$ & $3L^{[i|}_{\ \ m} V^{m|jk]}$ &
       $k_m\left(n^{i} V^{mjk}+
       n^{j} V^{mki}+n^{k} V^{mij}\right)$ \\
\hline $4$ & ${}^\bot V^{ijkl}$ & $4L^{[i|}_{\ \ m} V^{m|jkl]}$ & 
       $k_m\left(n^{i} V^{mjkl}-
       n^{j} V^{mkli}+ n^k V^{mlij}-
       n^{l} V^{mijk}\right)$ \\ 
\hline
\end{tabular}
\end{center}
An analogous scheme is valid for the corresponding densities.

\end{document}